\documentstyle{pss}
\input epsf.tex
\begin{document}

\title{ Vertical and Diagonal Stripes in the Extended \\ Hubbard Model }

\author{{\sc M. Raczkowski}\footnote{) Corresponding author; e-mail: 
marcin@alphetna.if.uj.edu.pl } ) (a), 
  {\sc B. Normand} (b), and {\sc A. M. Ole\'s} (a) }
\address{(a) Marian Smoluchowski Institute of Physics, Jagellonian 
 University,\\ 
 Reymonta 4, PL-30059 Krak\'ow, Poland \\
(b) D\'epartment de Physique, Universit\'e de Fribourg, 
 CH-1700 Fribourg, \\ Switzerland}
\submitted{May 21, 2002} 
\maketitle
\hspace{9mm} 

\begin{abstract}
We extend previous real--space Hartree--Fock studies of static stripe 
stability to determine the phase diagram of the Hubbard model with 
anisotropic nearest--neighbor hopping $t$, by varying the on--site 
Coulomb repulsion $U$ and investigating locally stable structures for 
representative hole doping levels $x=1/8$ and $x=1/6$. We also report 
the changes in stability of these stripes in the extended Hubbard model 
due to next--neighbor hopping $t'$ and to a nearest--neighbor Coulomb 
interaction $V$.  
\end{abstract}

Charge localization and the tendency of doped holes towards 
self--organization into striped patterns, observed in high-$T_c$
superconductors, is one of the most interesting current topics in the 
physics of strongly correlated electron systems \cite{Acta}. The 
stripe instability was predicted on the basis of mean--field calculations
before their experimental confirmation, in both three--band \cite{Zaa1} 
and one--band Hubbard models \cite{Inui}. These calculations yielded 
solutions with a phase separation which is manifest in the formation 
nonmagnetic lines of holes, one--dimensional domain walls or stripes, 
which separate antiferromagnetic (AF) domains of opposite phases. 
These phenomena are the most pronounced in 
La$_{1.6-x}$Nd$_{0.4}$Sr$_x$CuO$_4$ around hole doping $x=1/8$ \cite{Cu}, 
where the stripes are aligned along the two lattice directions $x$ and 
$y$, to which we refer as horizontal stripes (HS) or vertical stripes 
(VS). This is in contrast to the diagonal stripes (DS) inferred in the 
insulating nickelates La$_{2-x}$Sr$_x$NiO$_{4+y}$. However, although the 
multiband Hartree-Fock (HF) calculations of Zaanen and Littlewood 
\cite{Zaa2} are consistent with the observation of filled stripes in 
nickelates, by which is meant one doped hole per stripe site, the HF 
approximation does not predict the half--filled stripes (one hole every 
two atoms) observed in cuprates \cite{Zaa3}. In addition, charge 
transport in idealized stripes is not possible for the filled case. Both 
of these considerations indicate that it is necessary to go beyond the 
HF treatment of stripes by including local electron correlations which 
further influence the charge and spin distributions. However, significant 
qualitative statements remain possible within the framework of 
unrestricted HF. 
 
In this paper we attempt a systematic investigation of the properties and 
relative stability of filled VS and DS. We use the extended single--band 
Hubbard model, which is widely accepted as the generic model for a 
microscopic description of cuprate and nickelate systems,
\begin{equation}
H=-\sum_{ij\sigma}t^{}_{ij}c^{\dag}_{i\sigma}c^{}_{j\sigma} + 
  U\sum_{i}n^{}_{i\uparrow}n^{}_{i\downarrow} + 
  V\sum_{\langle ij\rangle}n^{}_in^{}_j,
\label{eq:h}
\end{equation}
where the operator $c^{\dag}_{i\sigma}$ $(c^{}_{j\sigma})$ creates 
(annihilates) an electron with spin $\sigma$ on lattice site $i$ ($j$), 
and $n_i=c^{\dag}_{i  \uparrow}c^{}_{i  \uparrow}
  +c^{\dag}_{i\downarrow}c^{}_{i\downarrow}$
gives the electron density. The hopping $t_{ij}$ is $t$ for nearest 
neighbors and $t'$ for second--neighbor sites $i$ and $j$, while the 
on--site and nearest--neighbor Coulomb interactions are respectively $U$ 
and $V$. The model is solved self-consistently in real space within the 
HF approximation, where the interactions are decoupled into products of 
one--particle terms, and we focus on the representative hole doping 
levels $x=1/8$ and $x=1/6$. We do not consider noncollinear spin 
configurations, and use the most straightforward version of the HF method 
with a product of two separate Slater determinants for up and down spins, 
whence $n_{i\uparrow}n_{i\downarrow}\simeq 
  n_{i\uparrow}\langle n_{i\downarrow}\rangle 
 +\langle n_{i\uparrow}\rangle n_{i\downarrow} 
 -\langle n_{i\uparrow}\rangle\langle n_{i\downarrow}\rangle$. 
A similar decoupling is performed for the nearest--neighbor Coulomb 
interaction. Calculations were performed on $12\times 12$ ($16\times 16$) 
clusters for $x=1/6$ ($x=1/8$) with periodic boundary conditions, and we 
obtain stable stripe structures with AF domains of width five atoms for 
$x=1/6$ and seven atoms for $x=1/8$. 
\begin{figure}[t!]
\begin{tindent}
\epsfxsize=12cm
\epsffile{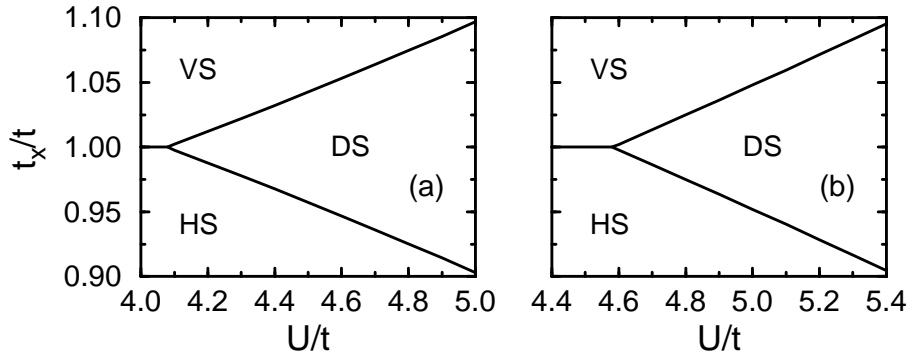}
\end{tindent}
\caption{
Phase diagrams for stable stripe structures obtained in the anisotropic 
Hubbard model ($t'=0$, $V=0$) for doping $x=1/8$ (a) and $x=1/6$ (b). }
\label{fig1}
\end{figure}
\begin{table}[b!]
\caption{ Site--normalized local charge density $n_{{\rm h}i} = 
\langle 1-(n_{i\uparrow} + n_{i\downarrow})\rangle$, local magnetization 
density $m_i^z={1\over 2}|\langle n_{i\uparrow}-n_{i\downarrow}\rangle|$, 
and kinetic energy contributions $E_{\rm K}^{x_i}$ and $E_{\rm K}^{y_i}$ 
on bonds between inequivalent atoms, all labeled by decreasing doped hole 
density in the $x$--direction, for VS (left) and DS (right) in the 
isotropic Hubbard model ($t_x=t_y$, $t'=0$, $V=0$) on a $16\times 16$ 
cluster with $U=5$ and $x=1/8$. In parenthesis are given the values of 
$n_{{\rm h}i}$ for the extended hopping model with $t'/t=-0.1$. }
\vskip 0.3cm
\begin{tabular}{ccccccccc}
\hline
$n^{}_{{\rm h}i}$ & $m_i^z$ & $E_{\rm K}^{x_i}/t$ & $E_{\rm K}^{y_i}/t$ & & 
$n^{}_{{\rm h}i}$ & $m_i^z$ & $E_{\rm K}^{x_i}/t$ & $E_{\rm K}^{y_i}/t$ \\
\hline
 0.364 (0.373) &0.000  &  $-$0.844 &  $-$0.643 & 
 &  0.388 (0.400) &0.000  &  $-$0.722 &  $-$0.722  \\
 0.234 (0.235) &  0.222 &  $-$0.662 &  $-$0.632 & 
 &  0.193 (0.195) &  0.262 &  $-$0.624 &  $-$0.722  \\
 0.067 (0.062) &  0.348 &  $-$0.600 &  $-$0.612 & 
 &  0.070 (0.067) &  0.352 &  $-$0.606 &  $-$0.624  \\
 0.014 (0.013) &  0.381 &  $-$0.595 &  $-$0.597 & 
 &  0.032 (0.029) &  0.372 &  $-$0.596 &  $-$0.606  \\
 0.006 (0.006) &  0.384 &  $-$0.595 &  $-$0.593 & 
 &  0.020 (0.019) &  0.380 &  $-$0.596 &  $-$0.596  \\
\hline
\end{tabular}
\label{tab1}
\end{table}

We begin by setting $t'=0$ and $V=0$. The phase diagrams shown in 
Fig.~\ref{fig1} were determined by varying $U$ and the ratio $t_x/t_y$ 
of the nearest--neighbor hoppings in the $x$-- and $y$--directions, 
while maintaining constant $t = {\textstyle \frac{1}{2}} (t_x + t_y)$. 
We observe the generic crossover from VS to DS with increasing 
Coulomb interaction reported in early HF studies \cite{Inui}. The 
transition from VS to DS appears in the isotropic case at $U/t\simeq 4.1$ 
for $x=1/8$ [Fig.~\ref{fig1}(a)], and at the higher value $U/t\simeq 4.6$ 
for $x=1/6$ [Fig.~\ref{fig1}(b)]. The results have a simple physical 
interpretation. Stripe phases occur as a compromise between on the one 
hand the AF interactions among magnetic ions and the Coulomb interactions 
which favor charge localization, and on the other the kinetic energy of 
doped holes which favors charge delocalization. Further, previous HF 
studies have clarified that the largest kinetic energy gains are obtained 
due to hopping perpendicular to the stripes \cite{Zaa3}. These features are 
seen in Tables 1 and 2. For VS one finds a large anisotropy in the values 
of the kinetic energies, $E_{\rm K}^{x_i}$ and $E_{\rm K}^{y_i}$, projected 
on the bonds in the $x$-- and $y$--directions, which becomes especially 
pronounced beside the stripes, and is strongly reinforced by the hopping 
anisotropy \cite{Bruce}. Thus, VS always lie along the direction of 
weaker hopping amplitude in the anisotropic model (Fig.~\ref{fig1}). 
\begin{figure}[t!]
\begin{tindent}
\epsfxsize=12cm
\epsffile{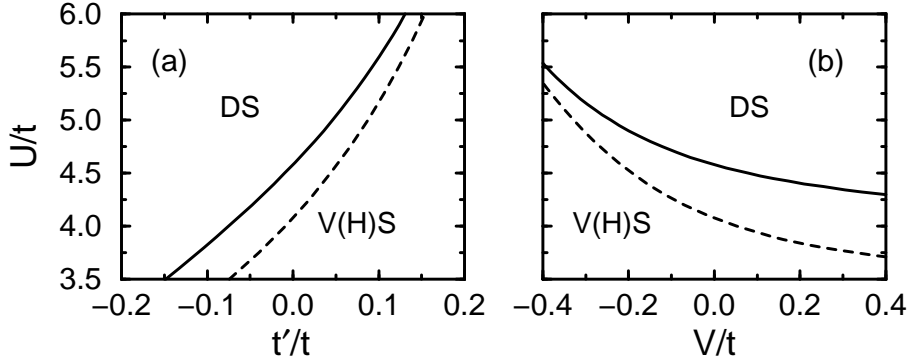}
\end{tindent}
\caption{
Phase boundaries for stripes in the isotropic extended Hubbard model with 
(a) $V=0$ and (b) $t'=0$, for dopings $x=1/8$ (dashed line) and $x=1/6$ 
(solid line). }
\label{fig2}
\end{figure}
\begin{table}[b!]
\caption{ Site--normalized ground--state energy $E_{\rm tot}$, kinetic 
energy $(E_{\rm K}^{x}, E_{\rm K}^{y}, E_{\rm K}^{x-y}, E_{\rm K}^{x+y})$, 
and potential energy $(E_U, E_V)$ components of VS (rows 1-5) and DS
(rows 6-10) in the isotropic extended Hubbard model on a $16\times 16$ 
cluster with $U=5$ and $x=1/8$. The DS is oriented along the direction 
$x-y$.  }
\vskip .3cm
\begin{tindent}
\begin{tabular}{rrccrrcrc}
\hline
$t'/t$ & $V/t$ & $E_{\rm K}^x/t$ & $E_{\rm K}^y/t$  & $E_{\rm K}^{x-y}/t$ & 
$E_{\rm K}^{x+y}/t$ & $E^{}_U/t$ & $E^{}_V/t$ &  $E^{}_{\rm tot}/t$ \\
\hline
0.0&  0.0  & $-$0.6753& $-$0.6147 & 0.0000& 0.0000  & 0.4900 & 
  0.0000  & $-$0.8000  \\
$-$0.1 &  0.0  & $-$0.6838& $-$0.5977 &0.0097  &  0.0097  & 0.4821 & 
  0.0000  & $-$0.7800  \\
0.1 &  0.0  & $-$0.6660& $-$0.6300 &  $-$0.0110 & $-$0.0110 & 0.4968 & 
  0.0000  & $-$0.8212  \\
0.0  &$-$0.4 & $-$0.6655& $-$0.6083& 0.0000& 0.0000  & 0.4749 & 
 $-$0.6251 & $-$1.4240  \\
0.0&  0.4 & $-$0.6838& $-$0.6214& 0.0000& 0.0000  & 0.5063 & 
0.6207  & $-$0.1782  \\  
\hline 
0.0&  0.0  & $-$0.6368& $-$0.6368 & 0.0000& 0.0000  & 0.4696 & 
  0.0000  & $-$0.8040  \\
$-$0.1 &  0.0  & $-$0.6309& $-$0.6309 & 0.0000&  0.0136  & 0.4587  &
  0.0000  & $-$0.7895  \\
0.1&  0.0  & $-$0.6417& $-$0.6417 & 0.0000& $-$0.0178& 0.4802 &  
  0.0000  & $-$0.8210  \\
0.0  &$-$0.4 & $-$0.6319& $-$0.6319& 0.0000& 0.0000  & 0.4602 & 
 $-$0.6193 & $-$1.4229  \\ 
0.0  & 0.4& $-$0.6412& $-$0.6412& 0.0000& 0.0000  & 0.4789 &
  0.6171  & $-$0.1864 \\
\hline
\end{tabular}
\end{tindent}
\label{tab2}
\end{table}

The kinetic energies in Table~\ref{tab2} show further that VS are more 
favorable for charge dynamics. This result, which is not immediately 
obvious, is due to the greater overall width of the stripe (Table~1), 
indicating that stripe fluctuations occur more readily in this geometry.
This explains their stability at small $U$ where the consequent cost in 
potential energy $E_U$ becomes less relevant. By contrast, DS have 
narrower stripes with larger magetization densities $m^z_i$ 
(Table~\ref{tab1}), meaning a lower net double occupancy and hence a 
more favorable $E_U$ (Table~\ref{tab2}). The transition from VS to DS 
with increasing $U$ is thus clarified. 

We have also considered the effect of a next--neighbor hopping $t'$ on
the relative stability of V(H)S and DS. Fig.~\ref{fig2}(a) shows that a 
negative $t'$ ($t'/t<0$), as obtained for the realistic parameters of 
high--$T_c$ superconductors, stabilizes DS \cite{nk2}, whereas a positive 
$t'$ favors VS, within the parameter range where $t'$ does not drive a 
stripe melting \cite{nk2}. The explanation is contained in Table~\ref{tab2}: 
negative $t'$ gives a positive kinetic energy contribution, which is much 
more readily minimized by the DS charge configuration. One observes 
further that positive $t'$ reduces the anisotropy between kinetic--energy 
gains in the $x$-- and $y$--directions for VS, and makes their sum more 
favorable, while negative $t'$ has the opposite effect. For DS the total 
kinetic energy also follows the same trend. The explanation for these 
results can be found in the reinforcement of stripe order by negative 
$t'$ (values in parenthesis in Table~1), which suppresses the hopping 
contributions, and its smearing out by positive $t'$ where hopping is 
enhanced. These trends agree with the earlier finding within the
dynamical mean field theory that VS are destabilized by kink fluctuations
\cite{dmft}. However, this stripe (dis)ordering tendency also leads to 
a considerably greater change in the Coulomb energy $E_U$ for DS than 
for VS (Table~\ref{tab2}), best seen in the charge--density alterations 
within the stripes (Table~\ref{tab1}), which contributes significantly 
to the predominance of DS for negative $t'$.
 
Finally, we investigate the changes in stripe stability due to repulsive 
($V>0$) and attractive ($V<0$) nearest--neighbor Coulomb interactions, 
which give the phase boundaries between VS and DS shown in 
Fig.~\ref{fig2}(b). Attractive $V$ enhances VS stability, while 
repulsive $V$ favors DS. The tendency towards VS formation at $V<0$  
is due primarily to their much higher charge densities on 
nearest-neighbor sites along the stripe, a situation which is avoided 
by DS (Table~\ref{tab2}). While this is also the leading mechanism for 
VS suppression at $V>0$, the asymmetry of the curve in Fig.~\ref{fig2}(b) 
arises from the fact that the lower $U$ values at the transition favor 
the higher kinetic energy contributions available for VS (above). 
  
In summary, we have shown that a competition between vertical (horizontal) 
and diagonal stripes dominates the behavior of the charge structures 
formed on doping the Hubbard model in the physically interesting regime 
of $3.5\le U/t\le 6$ within the HF approximation. The detailed charge 
distribution and the stripe ordering depend on the ratio $U/t$, on the 
value of the next--neighbor hopping $t'$, and on the nearest--neighbor 
Coulomb interaction $V$. Both repulsive $V$ and negative $t'$, which
correspond to the realistic parameters of high--$T_c$ superconductors,
act to reduce the relative stability of vertical (horizontal) stripes. 

{\it Acknowledgments.\/}
We thank R. Micnas and J. Zaanen for valuable discussions.
This work was financially supported by the Polish State Committee 
of Scientific Research (KBN), Project No. 5~P03B~055~20, and by the 
Swiss National Science Foundation.

\end{document}